\documentclass[aps,prl,reprint,twocolumn,amsmath,amssymb,showpacs,superscriptaddress,floatfix,longbibliography]{revtex4-1}

\usepackage{color}
\usepackage{textcomp} 
\usepackage{mathrsfs,amsmath}
\usepackage{graphicx}
\graphicspath{{Figures/}} 
\usepackage{lipsum}
\usepackage{dcolumn}
\usepackage[mathlines]{lineno}
\usepackage{hyperref}
\usepackage{xcolor}
\hypersetup{
    colorlinks,
    linkcolor={red!50!black},
    citecolor={blue!50!black},
    urlcolor={blue!80!black}
}
\usepackage{bm}
\usepackage{epstopdf}
\usepackage{soul}
\usepackage{epsfig}
\usepackage{bbold}
\usepackage{braket}
\usepackage{physics}
\usepackage{gensymb}
\usepackage{amsmath,amssymb}
\usepackage{bigints} %integral symbol of differnet sizes

\begin{document}

\title{Generating optical angular momentum through wavefront curvature}% Force line breaks with \\
\author{Kayn~A.~Forbes}
\email{K.Forbes@uea.ac.uk}

\affiliation{School of Chemistry, University of East Anglia, Norwich Research Park, Norwich NR4 7TJ, United Kingdom}

\author{Vittorio~Aita}

\affiliation{Department of Physics and London Centre for Nanotechnology, King's College London, Strand, London WC2R 2LS, UK}

\author{Anatoly~V.~Zayats}

\affiliation{Department of Physics and London Centre for Nanotechnology, King's College London, Strand, London WC2R 2LS, UK}

\begin{abstract}
Recent developments in the understanding of optical angular momentum have resulted in many demonstrations of unusual optical phenomena, such as optical beams with orbital angular momentum and transverse spinning light. Here we detail novel contributions to spin and orbital angular momentum generated by the gradient of wavefront curvature that becomes relevant in strongly focused beams of light. While circularly polarized beams are shown to develop helicity-dependent transverse spin, a linearly polarized Gaussian beam produces longitudinal spin and orbital angular momenta in the focal region, even if lacking both of these before focusing. 
Analytical treatment of a nonparaxial electromagnetic field, validated with vectorial diffraction modelling, shows that the terms related to higher orders of a paraxial parameter are responsible for the appearance of non-trivial angular momenta. The obtained dependences relate these quantities to the gradient of the wavefront curvature, showing how it can be used as a novel degree of freedom for applications in optical manipulation and light-matter interactions at subwavelength scales, enabling angular momentum transfer even from a simple Gaussian beam with linear polarization.
\end{abstract}

\maketitle

\section{Introduction}

%The story of light's momentum in free space for most of history was relatively short. 
Light with a wave vector $\mathbf{k}$ has a linear momentum along the direction of propagation of $\hbar\mathbf{k}$ per photon and an angular momentum of $\sigma\hbar\mathbf{\hat{k}}$, if elliptically polarized with helicity $\sigma$~\cite{allen2016optical}. Light beams with an azimuthal phase dependence ($\text{e}^{i\ell\phi}$), which are characterized by a topological charge $\ell \in \mathbb{Z}$, are often called vortex beams~\cite{allen1992orbital}. The realization that optical vortices could carry a well-defined orbital angular momentum of $\ell\hbar$ per photon, which is aligned along the propagation direction, resulted in many important ideas in optics~\cite{andrews2012angular, shen2019optical} and emergence of the field of structured light~\cite{gbur2016singular, rosales2018review, forbes2021structured, he2022towards}. 

Some of the most remarkable properties of structured light manifest themselves at subwavelength scales, where strong spatial confinement of the electromagnetic field can lead to the transverse spin angular momentum~\cite{bliokh2015transverse, aiello2015transverse, shi2023advances}, the longitudinal spin angular momentum of a linearly polarized optical vortex beam~\cite{yu2018orbit, forbes2021measures, forbes2024orbit, wu2024controllable}, and the polarization-independent optical helicity of vortex beams~\cite{wozniak2019interaction, forbes2021measures, forbes2022optical, green2023optical, forbes2023customized}. Under confinement, whether it be evanescent surface waves or tightly focused laser beams, electromagnetic fields in nano-optics are non-paraxial, possessing components in all three dimensions~\cite {novotny2012principles, alonso2023geometric}. It is the component along the direction of propagation--longitudinal field--which is predominantly responsible for such an extraordinary departure from the paraxial behaviour of light. 

In this work, we demonstrate counter-intuitive properties of nonparaxial light in a strong focusing regime: longitudinal spin and orbital angular momentum of a linearly polarized Gaussian beam and a helicity-dependent transverse spin momentum of circularly polarized beams. We show the origin of these novel contributions to optical angular momenta is the gradient of the wavefront curvature present in focused Gaussian laser beam modes. The results highlight that focusing an input beam with zero angular momentum would enable the transfer of both optical spin and orbital angular momenta to particles surrounding the focal plane for applications in tweezing and manipulation. One may foresee in particular the generation of spin from `spinless' beams to have important implications for spectroscopic applications in atomic and magneto-optics.

\section{Analytical description of a focused beam}

A paraxial beam is defined as one whose electromagnetic field is fully transverse to the propagation direction. This condition is typically realized for beams with the waist much bigger than the wavelength ($w_0\gg\lambda$). The electric field of a paraxial monochromatic Laguerre-Gaussian (LG) beam propagating along $z$ is 

\begin{align}
    \mathbf{E}^\text{T0}_{\text{LG}} = \bigl( \alpha \mathbf{\hat{x}} + \beta \mathbf{\hat{y}} \bigr) u_{\text{LG}}^{\ell,p}\,,
    \label{eq:1}
\end{align}
where $u_{\text{LG}}^{\ell,p}$ is~\cite{andrews2012angular}
\begin{align}
    u_{\text{LG}}^{\ell,p} &=  \nonumber \sqrt{\frac{2p!}{{\pi w_{0}^2}(p+|\ell|)!}}\frac{w_0}{w[z]}
\Biggr(\frac{\sqrt{2}r}{w[z]}\Biggr)^{|\ell|}
\\ & \nonumber \times L_{p}^{|\ell|}\Biggr[\frac{2r^2}{w^2[z]}\Biggr] \text{exp}(-r^2/w^2[z])
\\ &  \nonumber  \times \text{exp}i(kz+\ell\phi+kr^2/2R[z] -\omega t
\\ & - (2p + |\ell| + 1)\zeta [z])\,.
    \label{eq:2}
\end{align}
In the above $\alpha$ and $\beta$ are the (generally complex) Jones vector coefficients: $\abs{\alpha}^2+\abs{\beta}^2=1$, $\ell \in \mathbb{Z}$ and   $p \in \mathbb{Z^+}$  are the topological charge and radial index of the LG beam, respectively, $R[z]$ is the wavefront curvature, $\zeta[z]$ is the Gouy phase, $L_{p}^{|\ell|}\bigr[\frac{2r^2}{w^2[z]}\bigr]$ is the generalized Laguerre polynomial, square brackets are reserved for functional dependencies, all other symbols have their usual meaning, and the notation `T0' in Eq.~(1) will be explained below. The fundamental Gaussian mode can be constructed from the LG modes by setting $\ell =0, p=0$ in all derived expressions.

It is known that paraxial beams fail to satisfy Maxwell's equations~\cite{lax1975maxwell}. For example, Eq.~\eqref{eq:1} is clearly not divergence-free as required by the Gauss's law: $\nabla \cdot \mathbf{E}^\text{T0}_{\text{LG}} \neq 0$. Nonetheless, under paraxial conditions, Eq.~\eqref{eq:1} provides a good description. Beams which are strongly focused become nonparaxial, acquiring significant longitudinal electromagnetic field components. Under such circumstances, paraxial descriptions of the electromagnetic fields (Eq.~\eqref{eq:1}) clearly become unsatisfactory 
and alternative methods either analytical or numerical should be employed~\cite{peatross2017vector}. 
%Thanks to the strict boundary conditions required by numerical techniques based on diffraction integrals, these methods can provide closed-form solutions with strong quantitative accuracy. Analytical approaches on the other hand give much clearer physical insight into the physics of focused beams, explicitly showing how the multitude of higher-order electromagnetic field components contribute to optical properties like energy, momentum, angular momenta, etc. 

Under focusing, higher-order corrections to paraxial electromagnetic fields should be taken into account, which are generated proportionally to the so-called paraxial parameter. For Gaussian-type beams, it is $1/kw$. The paraxial field, e.g. that of Eq.~\eqref{eq:1}, is hence termed the zeroth-order transverse field T0 with respect to $1/kw$. 

The first non-paraxial correction is the first-order longitudinal field `L1', followed by the second-order transverse field `T2', and so on. Longitudinal (transverse) components are always odd(even)-order in the paraxial parameter. The analytical method of describing a non-paraxial beam involves taking the zeroth-order description of the electromagnetic field of Eq.~\eqref{eq:1}, which has a nonzero divergence, and using Maxwell's equations in an iterative process to generate the higher-order terms. 
The unknown term `L1' is obtained imposing the Gauss's law on $ \mathbf{E}^\text{T0} + \mathbf{E}^\text{L1}$:

\begin{align}
    \nabla \cdot \mathbf{E}^\text{T0 + L1} = \nabla_{\perp} \cdot \mathbf{E}^\text{T0} + \frac{\partial }{\partial z}E^\text{L1}_z = 0\,,
    \label{eq:3}
\end{align}
therefore:
\begin{align}
    E^\text{L1}_z =-\int \nabla_{\perp} \cdot \mathbf{E}^\text{T0} {\partial z}
    \label{eq:4}
\end{align}

In order to secure an analytical result for $E_z$, it is at this point the approximation is made that the variation in $z$ is dominated by the $\text{e}^{ikz}$ phase factor, which leads to~\cite{adams2018optics}
\begin{align}
    E^\text{L1}_z \approx \frac{i}{k} \nabla_{\perp} \cdot \mathbf{E}^\text{T0}\,.
    \label{eq:5}
\end{align}
This shows that the first-order longitudinal electric field component $E^\text{L1}_z$ is directly proportional to the transverse gradient of the zeroth-order transverse field. Applying Gauss's law to the obtained field, obviously results in a zero divergence ($\nabla \cdot (\mathbf{E}^\text{T0}_{\text{LG}} + \mathbf{\hat{z}}{E}^\text{L1}_{\text{LG}}) = 0$), as required by the Maxwell's equations. 
Continuing the iteration, the field $\mathbf{E}^\text{T0}_{\text{LG}} + \mathbf{E}^\text{L1}_{\text{LG}}$ is inserted into the Faraday's Law to generate the magnetic field, up to the second order in $1/kw$ ($\mathbf{B}^\text{T0}_{\text{LG}} + \mathbf{B}^\text{L1}_{\text{LG}} + \mathbf{B}^\text{T2}_{\text{LG}}$). Finally,  $\mathbf{B}^\text{T0 + L1}_{\text{LG}}$ is plugged in the Maxwell-Ampere law, yielding the second-order transverse electric field  $\mathbf{E}^\text{T0}_{\text{LG}} + \mathbf{E}^\text{L1}_{\text{LG}} + \mathbf{E}^\text{T2}_{\text{LG}}$. 
With a single application of the iteration, both fields are obtained up to their second-order term in $1/kw$. To ensure the correct description, it is essential that the electric and magnetic fields are derived up to the same order. The resulting expression for the electric field is

\begin{widetext}
\begin{align}
    \mathbf{E}^\text{T0+L1+T2}_{\text{LG}} &=  u_{\text{LG}}^{\ell,p}\, \Bigl\{\overbrace{\alpha \mathbf{\hat{x}} + \beta \mathbf{y}}^{\text{T0}} + \overbrace{\mathbf{\hat{z}}\frac{i}{k} \Bigl[\alpha \bigl( \gamma \cos\phi - \frac{i\ell}{r} \sin\phi \bigr) + \beta \bigl( \gamma \sin\phi + \frac{i\ell}{r} \cos\phi \bigr) \Bigr]}^{\text{L1}} \nonumber \\
    &+ \frac{1}{k^2} \mathbf{\hat{x}} \Bigl[2\alpha\sin\phi\cos\phi\Bigl(\frac{i\ell}{r^2}-\frac{i\ell\gamma}{r}\Bigr) + \alpha\cos^2\phi\Bigl(\frac{\ell^2}{r^2}-\frac{\gamma}{r}\Bigr)-\alpha\sin^2\phi\bigl\{\gamma^{'} + \gamma^2\bigr\} \nonumber \\ & + \beta\sin\phi\cos\phi\Bigl(\gamma^{'} + \gamma^2 - \frac{\gamma}{r} + \frac{\ell^2}{r^2} \Bigr) + \beta \cos^2\phi \Bigl(\frac{i\ell\gamma}{r} - \frac{i\ell}{r^2} \Bigr) + \beta \sin^2\phi \Bigl(\frac{i\ell}{r^2} - \frac{i\ell\gamma}{r} \Bigr) \Bigr) \nonumber \\
    &+ \frac{1}{k^2} \mathbf{\hat{y}} \Bigl[\alpha\sin\phi\cos\phi\Bigl(\gamma^{'}+\gamma^2-\frac{\gamma}{r}+\frac{\ell^2}{r^2}\Bigr) + \alpha\cos^2\phi\Bigl(\frac{i\ell\gamma}{r} - \frac{i\ell}{r^2}\Bigr) + \alpha\sin^2\phi\Bigl(\frac{i\ell}{r^2}-\frac{i\ell\gamma}{r}\Bigr) \nonumber \\
    &+ 2\beta\sin\phi\cos\phi\Bigl(\frac{i\ell\gamma}{r} - \frac{i\ell}{r^2}\Bigr) -\beta\cos^2\phi\Bigl(\gamma^{'} + \gamma^{2}\Bigr) + \beta\sin^2\phi\Big(\frac{\ell^2}{r^2}-\frac{\gamma}{r}\Bigr)\Bigr]\Bigr) \,,
    \label{eq:6}
\end{align}
\end{widetext}
where 
\begin{align}
    \gamma\qty(\vb{r}) &= \frac{|\ell|}{r}-\frac{2r}{w^2}+ \frac{ikr}{R[z]} -\frac{4r}{w^2}\frac{L^{|\ell|+1}_{p-1}}{L^{|\ell|}_{p}}\,,
    \label{eq:7}
\end{align}
which comes from the derivative of the field over the radial coordinate $r$: $\frac{\partial}{\partial r} u_{\text{LG}}^{\ell,p} = \gamma u_{\text{LG}}^{\ell,p}$. Note that $\gamma^{'} = \frac{\partial}{\partial r} \gamma$ and for $p=0$, ${L^{|\ell|+1}_{p-1}}=0$.

The longitudinal and the higher-order transverse components of the field in Eq.~\ref{eq:6} are of order one and two in the paraxial parameter, respectively so that in a paraxial beam they are negligible compared to the zeroth order term T0. In tightly focused beams, as $w_0$ becomes comparable in size with $\lambda$, the paraxial parameter $1/kw\longrightarrow1$, meaning that the paraxial approximation is broken and the higher-order fields thus become appreciable with respect to the T0 term. By fixing the beam waist to wavelength ratio, Eq.~\eqref{eq:6} can model both a paraxial and non-paraxial LG mode of any order $(\ell,p)$.  

In the paraxial case (Eq.~\ref{eq:1}), the LG modes can have arbitrary two-dimensional (2D) states of polarization, defined by the Jones vector $\qty(\alpha,\beta)$. Beyond the paraxial approximation, approaching the focal plane, the polarization of the beam will be non-trivial and three-dimensional (3D)~\cite{alonso2023geometric}. Throughout this work, we refer to the state of polarization in the $(x,y)$-plane of the input $z$-propagating paraxial beam described by the Jones vector before focusing, i.e., the T0 field components, as the 2D state of polarization.  
%although the form of Eq.~\eqref{eq:6} is somewhat daunting, 
The full 3D polarization state of a focused, nonparaxial beam can be obtained from Eq.~\eqref{eq:6} by substituting in the values of $\qty(\alpha,\beta)$ that describe the 2D polarization state of the input paraxial beam. For example, with $\qty(1,0)$ or $\qty(0,1)$ for a 2D $x$-polarized or $y$-polarized LG beam, respectively, or with $\qty(1,\pm i)/\sqrt{2}$ for a 2D circularly polarized one.

%Having clarified the role of higher order terms in $1/kw$, it is the imaginary part of Eq.~\eqref{eq:7} that takes centre stage in this work. 
The imaginary part of Eq.~\eqref{eq:7} $\Im{ \gamma\qty(\vb{r})} = ikr/R[z]$ corresponds to the gradient of the wavefront curvature $R[z]$ and provides the counter-intuitive contributions to the angular momentum of focused beams we concentrate on in this work. Since all the quantities used in the following are proportional to $\qty(1/kw)^2$, a truncation of the nonparaxial electromagnetic field to the second order in the paraxial parameter is justified.

\section{Spin Angular Momentum}
The cycle-averaged electric spin angular momentum density for a monochromatic beam is given by~\cite{bliokh2013dual} 

\begin{align}
\mathbf{s_E} = \frac{\epsilon_0}{2} \Im\qty(\mathbf{E}^*\times\mathbf{E})\,, 
\label{eq:8}
\end{align}
which can be understood as the degree to which orthogonal components of the electric field are $\pi/2$ out of phase with one another. In other words, Eq.~\eqref{eq:8} quantifies the ellipticity of the polarization in a given plane.
By inserting Eq.~\eqref{eq:6} into Eq.~\eqref{eq:8}, it is possible to obtain the spin density up to the second order in the paraxial parameter and to analyse the contribution of each order to it separately. Assuming that the electric field is written as $\mathbf{E}^\text{T0} +\mathbf{E}^\text{L1}+ \mathbf{E}^\text{T2} $, the cross product in Eq.~\eqref{eq:8} will contain a term of order zero coming from T0$\cross$T0, terms of order one from T0$\cross$L1 and L1$\cross$T0, and terms of order two from T0$\cross$T2 and T2$\cross$T0. The other possible combinations would all be a $O\qty((1/kw)^2)$, so that Eq.~\eqref{eq:8} can be written as a sum of the following contributions:
\begin{widetext}
    \begin{align}
        \mathbf{s_E}^{\text{T0} \times \text{T0}} &= \frac{\epsilon_0}{2} \Im \bigl(\mathbf{E}^{*\text{T0}}_\text{LG} \times \mathbf{E}^{\text{T0}}_\text{LG}\bigr) \nonumber \\ 
        & = \mathbf{\hat{z}}\frac{\epsilon_0}{2}|u_{\text{LG}}^{\ell,p}|^2\,\Im\bigl(\alpha^{*}\beta - \alpha \beta^{*}\bigr)\,,\label{eq:9}\\
        \mathbf{s_E}^{\text{T0} \times \text{L1}} &= \frac{\epsilon_0}{2} \Im \bigl(\mathbf{E}^{*\text{T0}}_\text{LG} \times \mathbf{E}^{\text{L1}}_\text{LG} + \mathbf{E}^{*\text{L1}}_\text{LG} \times \mathbf{E}^{\text{T0}}_\text{LG} \bigr) \nonumber \\ 
        & = |u_{\text{LG}}^{\ell,p}|^2\,\Im\Bigl\{\frac{i\epsilon_0}{2k}\Bigl[ \mathbf{\hat{x}} \Bigl(\alpha\beta^* \bigl(\gamma\cos\phi-\frac{i\ell}{r}\sin\phi\bigr) + |\beta|^2\bigl(\gamma\sin\phi+\frac{i\ell}{r}\cos\phi\bigr) + \text{c.c.}\Bigr) \nonumber \\ 
        & - \mathbf{\hat{y}} \Bigl(\alpha^{*}\beta \bigl(\gamma\sin\phi+\frac{i\ell}{r}\cos\phi\bigr) + |\alpha|^2\bigl(\gamma\cos\phi-\frac{i\ell}{r}\sin\phi\bigr) + \text{c.c.}\Bigr) \Bigr]\Bigr\}\,, \label{eq:10}\\
        \mathbf{s_E}^{\text{T0} \times \text{T2}} &= \frac{\epsilon_0}{2} \Im \bigl(\mathbf{E}^{*\text{T0}}_\text{LG} \times \mathbf{E}^{\text{T2}}_\text{LG} + \mathbf{E}^{*\text{T2}}_\text{LG} \times \mathbf{E}^{\text{T0}}_\text{LG} \bigr) \nonumber \\ 
        & = \mathbf{\hat{z}}|u_{\text{LG}}^{\ell,p}|^2\,\Im\Bigl\{\frac{\epsilon_0}{2k^2}\Bigl[ \Bigl(|\alpha^2|\sin\phi\cos\phi\Bigl(\gamma^{'}+\gamma^2-\frac{\gamma}{r}+\frac{\ell^2}{r^2}\Bigr) + 2\alpha^*\beta\sin\phi\cos\phi\Bigl(\frac{i\ell\gamma}{r}-\frac{i\ell}{r^2}\Bigr) \nonumber \\
        &+ |\alpha^2|\cos^2\phi\Bigl(\frac{i\ell\gamma}{r}-\frac{i\ell}{r^2}\Bigr) + |\alpha^2|\sin^2\phi \Bigr\{\frac{i\ell}{r^2}-\frac{i\ell\gamma}{r}\Bigr) -\alpha^*\beta\cos^2\phi\bigl\{\gamma^{'}+\gamma^2\bigr\} + \alpha^*\beta\sin^2\phi\Bigl(\frac{\ell^2}{r^2}-\frac{\gamma}{r}\Bigr) -\text{c.c.} \Bigr) \nonumber \\  
        & - \Bigr(2\alpha\beta^*\sin\phi\cos\phi\Bigl(\frac{i\ell}{r^2}-\frac{i\ell\gamma}{r}\Bigr) + \alpha\beta^*\cos^2\phi\Bigl(\frac{\ell^2}{r^2}-\frac{\gamma}{r}\Bigr)-\alpha\beta^*\sin^2\phi\bigl\{\gamma^{'}+\gamma^2\bigr\} \nonumber \\ 
        & + |\beta^2|\sin\phi\cos\phi\Bigl(\gamma^{'}+\gamma^2-\frac{\gamma}{r}+\frac{\ell^2}{r^2}\Bigr) + |\beta^2|\cos^2\phi\Bigl(\frac{i\ell\gamma}{r}-\frac{i\ell}{r^2}\Bigr) + |\beta^2|\sin^2\phi\Bigl(\frac{i\ell}{r^2}-\frac{i\ell\gamma}{r}\Bigr) - \text{c.c.} \Bigr) \Bigr]\Bigr\}\,.\label{eq:11}
    \end{align}
\end{widetext}

\subsection{Longitudinal spin}
\begin{figure*}[ht]
    \includegraphics[width = \linewidth]{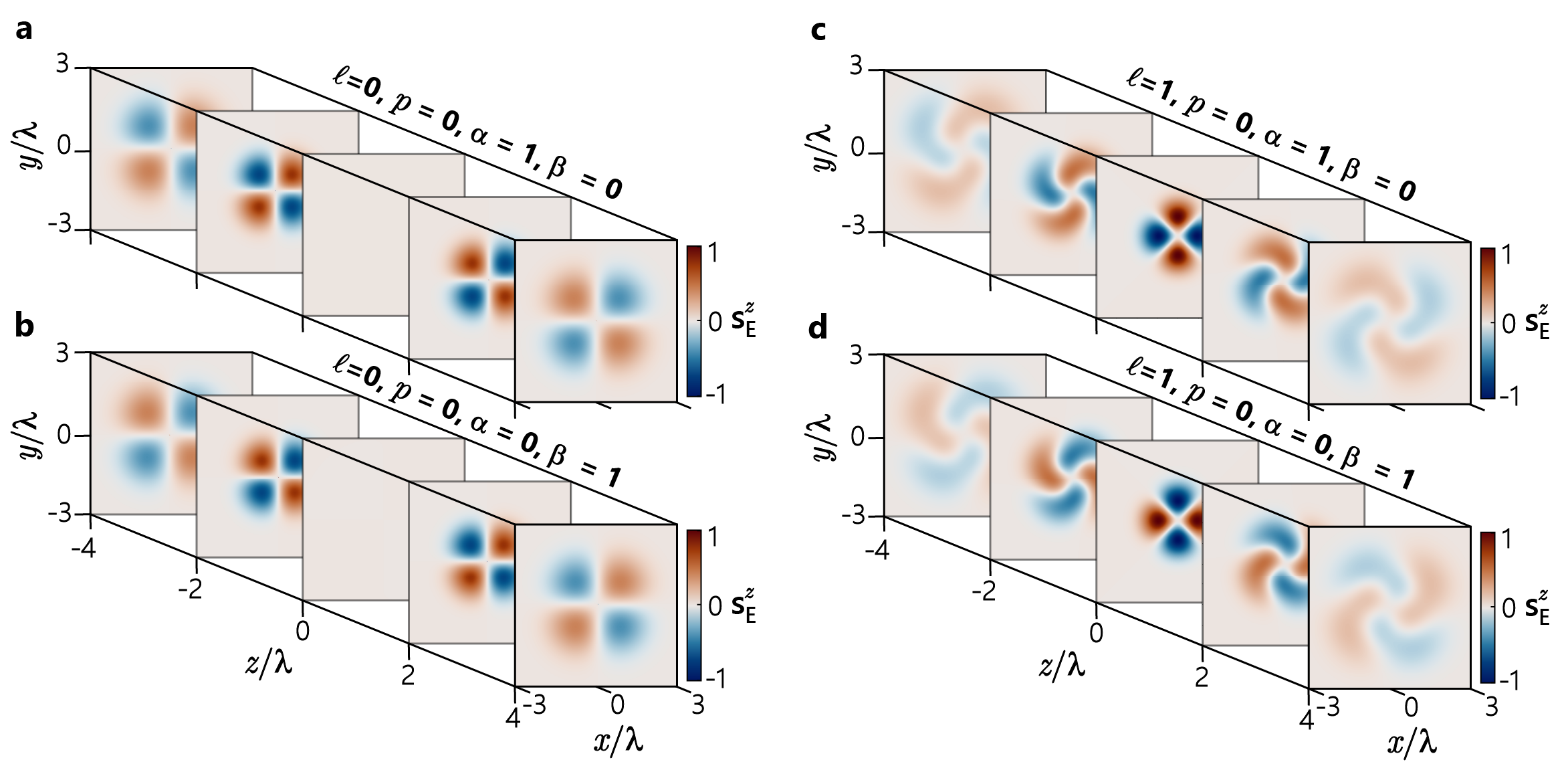}
    \caption{The dependence of longitudinal spin angular momentum density of 2D linearly polarized LG beams on the propagation distance$z$, simulated from Eq.~\eqref{eq:11} for the beam with $w_0 / \lambda = 1$. The beam quantum numbers are (a,b) $\ell = 0, p = 0$ and (c,d) $\ell = 1, p = 0$, and their polarization is chosen along the (a,c) $x$-axis, with $\alpha = 1$, $\beta = 0$ and the (b,d) $y$-axis, with $\alpha = 0$, $\beta = 1$. In the focal plane $z=0$, the spin angular momentum density (Eq.~\eqref{eq:13}) is zero for a fundamental Gaussian beam (a,b), however, around the focal plane, it is non-zero. The colour scale~\cite{crameri2020misuse} is common to all the plots, each normalized to the peak value of the set. }
    \label{fig:1}
\end{figure*}
The longitudinal component of the spin angular momentum density ($s_\mathrm{E}^z$) comes from the terms T0$\cross$T0 and T0$\cross$T2 (Eqs.~\eqref{eq:9},\eqref{eq:11}). In particular, the zeroth order contribution is always zero unless the input polarization state $\qty(\alpha,\beta)$ is complex, meaning that the beam has a nonzero ellipticity. This is the standard, well-known origin of spin angular momentum for plane waves and paraxial beams of light. The second-order term is instead present regardless of the input polarization. A linearly polarized LG beam hence becomes an interesting case study as the only nonzero term in the longitudinal spin density is the one coming from nonparaxiality (Eq.~\eqref{eq:11}).

The spatial distribution of the longitudinal spin angular momentum (Eq.\eqref{eq:11}) for a focused Gaussian beam has a counter-intuitive property: around the focal plane, but not at $z=0$, a nonzero spin angular momentum density is generated even for linearly polarized beams which do not carry helicity in the paraxial case (Fig.~\ref{fig:1}a,b). The spatial distribution is acutely sensitive to the orientation (the azimuth on the Poincaré sphere) of the 2D state of linear polarization, and rotates accordingly (Supplementary Fig.~\ref{fig:si_extra_cases}a,b). Each lobe of spin changes sign upon crossing of the focal plane. Additionally, beams carrying orbital angular momentum $\ell\neq0$ show a nonzero longitudinal spin momentum density even in the focal plane $z=0$ (Fig.~\ref{fig:1}c,d). Its spatial distribution is somewhat twisted around the focal plane and depends on the polarization orientation as per non-vortex beam, but also on the sign of $\ell$ (Supplementary Fig.~\ref{fig:si_extra_cases}c,d). 
\begin{figure*}[!ht]
    \includegraphics[width = \linewidth]{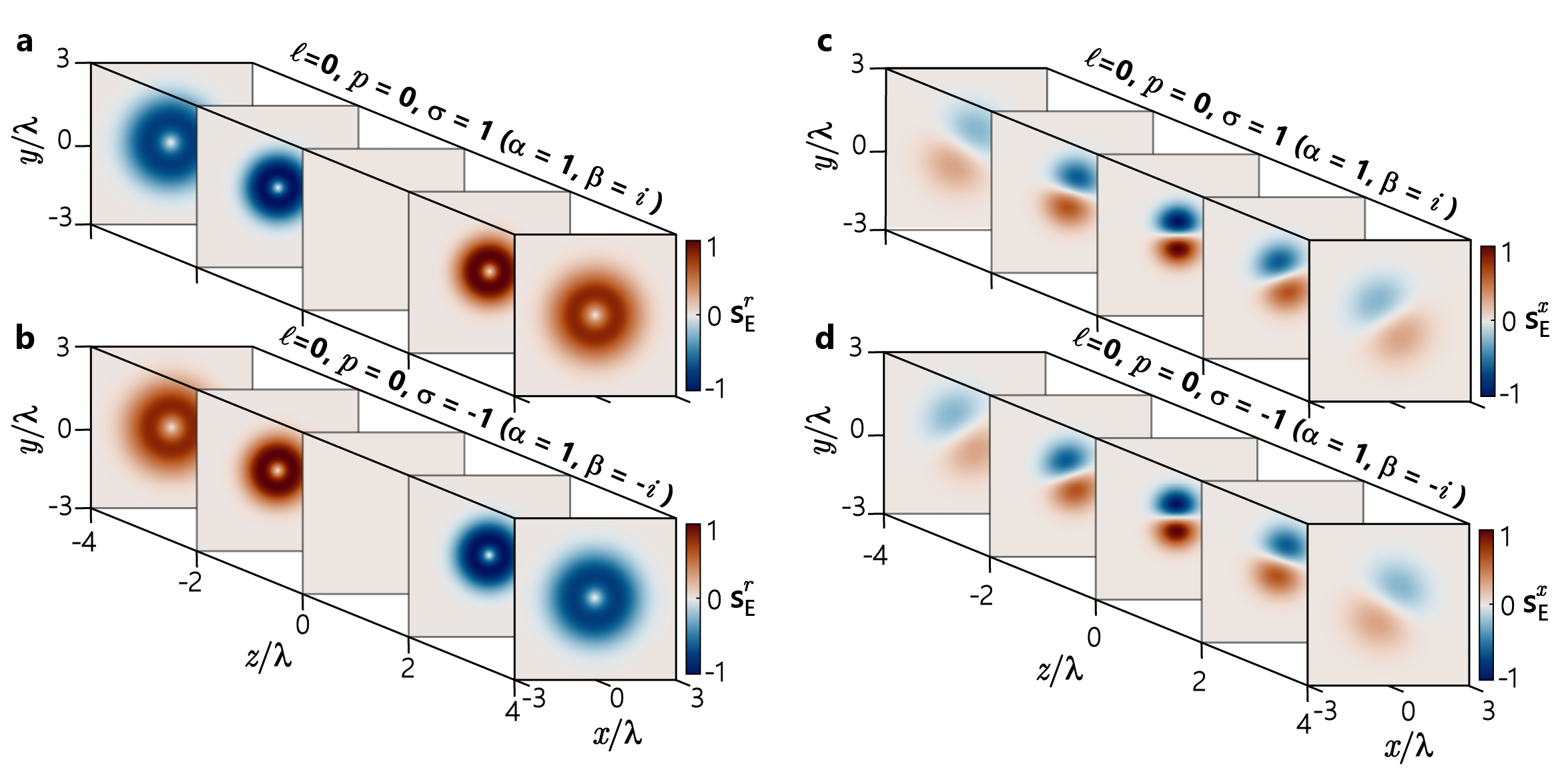}
    \caption{Components of the transverse spin momentum (a,b) $\mathbf{s_E}^{\text{r}}$ and (c,d) $\mathbf{s_E}^{\text{x}}$  (Eq.~\eqref{eq:14}), for a 2D-circularly polarized fundamental Gaussian beam of (a,c) left- ($\sigma = 1$) and (b,d) right- ($\sigma = -1$). The $y$ component is shown in Supplementary Fig.~\ref{fig:si_ycomps}a,b. The helicity-dependence is clearly visible with the spin taking the same sign as the helicity as it propagates away from the focal point. Each plot is normalized to the peak value of the set. The beam parameters are as in Fig.~1.}
    \label{fig:trans_comps_lg00_circ}
\end{figure*} 
The underlying physics of this behaviour can be ascertained from the analytical formula Eq.~\eqref{eq:11} when specific beam parameters are given as input. For $x$-polarized beams ($\alpha=1, \beta=0$), the spin angular momentum density is given by 
\begin{align}
\mathbf{s_E}^{\alpha = 1} &= \mathbf{\hat{z}}\frac{\epsilon_0}{k^2}\,|u_{\text{LG}}^{\ell,p}|^2\,\Biggl\{\frac{k\sin2\phi}{R[z]}\Biggl(|\ell|-\frac{2r^2}{w^2}-\frac{4r^2L^{|\ell|+1}_{p-1}}{w^2L^{|\ell|}_{p}}\Biggr) \nonumber \\ & + \cos2\phi\Biggl[\frac{\ell}{r}\Biggl(\frac{|\ell|}{r}-\frac{2r}{w^2}-\frac{4rL^{|\ell|+1}_{p-1}}{w^2L^{|\ell|}_{p}}\Biggr)-\frac{\ell}{r^2}\Biggr]\Biggr\}\,.
\label{eq:12}
\end{align}
Here, the first term in curly brackets proportional to $\sin2\phi$ depends on the wavefront curvature $R[z]$ and is thus responsible for the twisting effect observed for $z\neq0$ in Figure~\ref{fig:1}c,d. The terms with $\cos2\phi$ generate spin angular momentum density at all $z$ distances and bring the dependence on the sign of $\ell$.
In the simplest case of a linearly polarized fundamental Gaussian beam ($\ell=0,p=0$), Eq.~\eqref{eq:12} further simplifies to
\begin{align}
\mathbf{s_E}^{\alpha = 1}_{\ell=p=0} &=- \mathbf{\hat{z}}\frac{2\epsilon_0 r^2}{kw^2\,R[z]}\sin2\phi\,|u_{\text{LG}}^{0,0}|^2\,, 
\label{eq:13}
\end{align}
where the explicit definition of the wavefront curvature is $R[z] = (z^2 + z_R^2)/z$. Albeit the simplicity, this unexpected observation shows that a 2D linearly polarized Gaussian beam produces a nonzero spin angular momentum density at all $z$ but the focal plane. The origin of such a peculiar effect can be traced back to the behaviour of the wavefront curvature as a function of $z$, which diverges to $+\infty$ at both the focal plane and infinity, hence bringing the longitudinal spin density to zero. The strongest $s_{\mathrm{E}}^z$ is found at the edges of the Rayleigh range ($\pm z_R$), where the wavefront curvature drops to its minimum.

Similar results can be obtained for a $y$-polarized Gaussian beam ($\alpha = 0, \beta = 1$), showing that $\mathbf{s_E}^{\beta = 1} = - \mathbf{s_E}^{\alpha = 1}$ for any value of the quantum numbers $\ell$ and $p$ (cf. Figs~\ref{fig:1}a and b). Although the key Eqs.~\eqref{eq:9}-\eqref{eq:11} are completely general with respect to the state of 2D polarization, linearly polarized beams attract particular interest as their behaviour in the near field of the focal plane is substantially different from their far field, where they have zero spin, i.e. spin can be generated from 'spinless' beams by focusing. 

For 2D polarization states with a nonzero ellipticity, this analytical model can be used to calculate the second-order longitudinal spin density that, albeit nonzero, is shown to be proportional to $1/R[z]^2$, which is much smaller than what is obtained for linearly polarized beams (cf. Eq.~\eqref{eq:13}). Moreover, its identification in an experiment would be seemingly impossible when the leading-order longitudinal spin for a beam with ellipticity (Eq.~\eqref{eq:9}) would dominate other effects. This is different from linearly polarized beams, for which there is no first-order spin density.
%In summary, in this section we have shown how the gradient of the wavefront curvature of linearly polarized Laguerre-Gaussian beams generates longitudinal spin angular momentum density under tight focusing.

\subsection{Transverse spin}
The first nonzero order of transverse component of the spin angular momentum density in Eq.~\eqref{eq:8} comes from the T0$\cross$L1 term Eq.~\eqref{eq:10}, appearing at the order one in $1/kw$. The wavefront curvature contribution to the transverse spin Eq.~\eqref{eq:10} comes from the imaginary part of $\gamma$ ($\Im\gamma = kr/R[z]$) so that given the factor of $i$ at the front of Eq.~\eqref{eq:10} it can only contribute for input beams with complex parameters $\alpha,\beta$. Hence, for beams with linear 2D polarization ($\alpha,\beta\in\mathbb{R}$), the wavefront curvature does not contribute to the
transverse spin, as the terms dependent on the gradient of the wavefront curvature are real. On the contrary, a transverse spin density is generated by the gradient of the wavefront curvature for beams with a nonzero ellipticity in their 2D input polarization state (i.e., $\beta$ is imaginary). In particular, for circular polarization ($\alpha = 1/\sqrt{2}, \beta = i\sigma/\sqrt{2}$) of helicity $\sigma = \pm1$  the transverse spin density of Eq.~\eqref{eq:10} can be written as 

\begin{align}
\mathbf{s_E}^{\sigma,\perp}&= \frac{\epsilon_0}{2k}|u_{\text{LG}}^{\ell,p}|^2\,\Bigl[\mathbf{\hat{r}}\frac{\sigma k  r}{R[z]} \nonumber \\ 
& - \mathbf{\hat{\phi}}\Bigl(\frac{|\ell|}{r}-\frac{2r}{w^2}-\frac{4r}{w^2}\frac{L_{p-1}^{|\ell|+1}}{L_p^{|\ell|}}-\frac{\ell \sigma}{r}\Bigr)\Bigr]\,, 
\label{eq:14}
\end{align}
where only the $\mathbf{\hat{r}}$ component depends on the wavefront curvature. In the case of a circularly polarized Gaussian beam, this component of transverse spin is helicity-dependent(Fig.~\ref{fig:trans_comps_lg00_circ}a,b).This is in stark contrast to the helicity-independent transverse spin of evanescent waves ~\cite{bliokh2015transverse}~and has gone unnoticed in the focused beams explored to date ~\cite{bliokh2014extraordinary, eismann2021transverse}. The spin takes on the opposite sign to the helicity as the beam propagates towards the focal point, where the wavefront is converging, and the same sign as the helicity past the focal point, where the wavefront is diverging. It is worth noting that the dependence on the quantum numbers $\ell$ and $p$ is only encoded in the amplitude term $|u_{\text{LG}}^{\ell,p}|^2$, so that the observations made above for a Gaussian beam can be generalized to any LG beam.

For the previously examined case of longitudinal spin, the introduction of a nonzero topological charge causes a twist in the spin density along the $z$ direction. A similar result is obtained for the transverse spin density for beams of nonzero helicity, even if with a zero topological charge. This is best highlighted by presenting the transverse spin density Eq.~\eqref{eq:14} in Cartesian coordinates: 
\begin{align}
\mathbf{s_E}^{\sigma,x}&= \frac{\epsilon_0}{2k}|u_{\text{LG}}^{\ell,p}|^2\Bigl[\cos\phi\frac{\sigma k r}{R[z]} \nonumber \\ 
& + \sin\phi\Bigl(\frac{|\ell|}{r}-\frac{2r}{w^2}-\frac{4r}{w^2}\frac{L_{p-1}^{|\ell|+1}}{L_p^{|\ell|}}-\frac{\ell \sigma}{r}\Bigr)\Bigr]\,, 
\label{eq:15}
\end{align}
\begin{align}
\mathbf{s_E}^{\sigma,y}&= \frac{\epsilon_0}{2k}|u_{\text{LG}}^{\ell,p}|^2\Bigl[\sin\phi\frac{\sigma kr}{R[z]} \nonumber \\ 
& - \cos\phi\Bigl(\frac{|\ell|}{r}-\frac{2r}{w^2}-\frac{4r}{w^2}\frac{L_{p-1}^{|\ell|+1}}{L_p^{|\ell|}}-\frac{\ell \sigma}{r}\Bigr)\Bigr]\,,
\label{eq:16}
\end{align}
Starting from the case of null topological charge (Fig.~\ref{fig:trans_comps_lg00_circ}c,d), the results show that the $x$ component of the spin density does not depend on the sign of the helicity in the focal plane, while it is strongly influenced by $\sigma$ beyond this point. In particular, looking at the beam propagating along $z>0$, the spatial distribution rotates anticlockwise if $\sigma$ and $R[z]$ have the same sign and clockwise otherwise. Opposite behaviour for the same component of the beams with opposite helicity ($y$ components is observed (Supplementary Fig.~\ref{fig:si_ycomps}).

\begin{figure}[ht]
    \includegraphics[width = \linewidth]{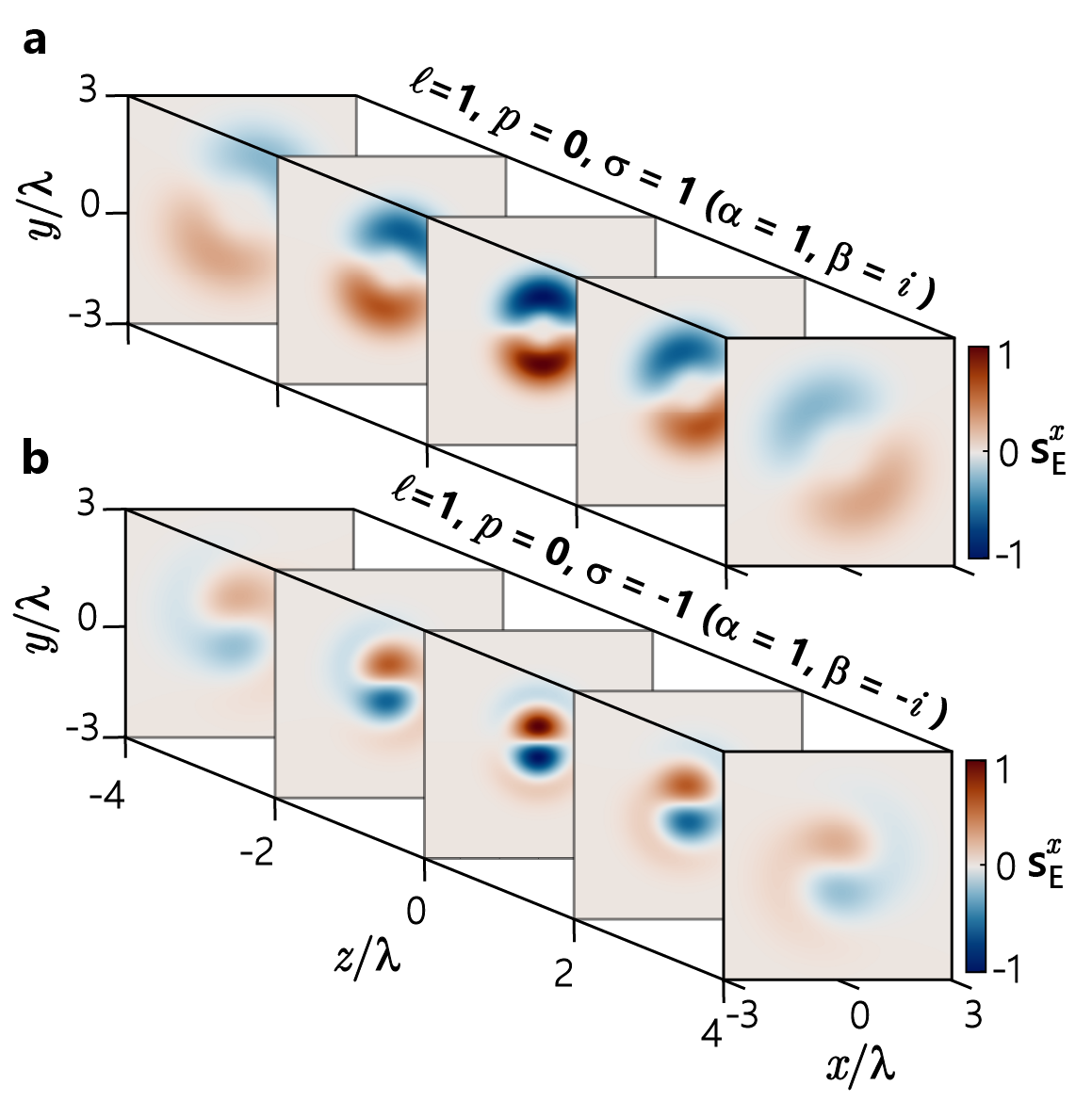}
    \caption{The dependence of the transverse spin density component $\mathbf{s_E}^{x}$ on a propgation distance $z$ for an LG beam of parameters $\ell = 1$, $p=0$ and circular polarization of (a) left- and (b) right- handedness. These result in the (a) ``parallel'' and (b) ``anti-parallel'' configurations of the spin-orbit interaction. The plot for $\mathbf{s_E}^{y}$ is shown in Supplementary Fig.~\ref{fig:si_ycomps}c,d}
    \label{fig:x_comp_lg10_circ}
\end{figure} 

More interesting is the case of circularly polarized LG beams with nonzero topological charge (Fig.~\ref{fig:x_comp_lg10_circ}). The tight focusing is shown to enable a spin-orbit coupling interaction, illustrated by the evident difference between the cases of $\ell$ and $\sigma$ having the same (parallel configuration) or opposite (antiparallel configuration) sign. In the former case--``parallel'' spin-orbit--the $x$ component of the spin density acquires a two-lobed shape featuring the same twisting mechanism highlighted for circular beams with no topological charge, although with a different spatial distribution  (Fig.~\ref{fig:x_comp_lg10_circ}a). In the ``anti-parallel'' configuration, the transverse spin density shows a 4-lobed distribution in the focal plane and a double branch spiral one beyond it (Fig.~\ref{fig:x_comp_lg10_circ}b). It is worth noting that the spin density spatial distribution undergoes a twisting along $z$ only in the parallel configuration

 Finally, whilst both the longitudinal spin densities of Laguerre-Gaussian and Gaussian beams (Eqs.~\eqref{eq:12} and \eqref{eq:13}, respectively) show the rotation in their spatial distribution due to the wavefront curvature for any value of $p$, the transverse spin momentum density (Eq.~\eqref{eq:14}) only rotates for $p=0$. The reason for this can be found in the azimuthal contribution of the Laguerre polynomials to the spin density, which is only zero for $p=0$, while for $p>0$, the second term in square brackets of Eqs.~\eqref{eq:15} and \eqref{eq:16} quickly overcomes the helicity-dependent terms responsible for the rotation~(Supplementary Fig.~\ref{fig:si_nonzeroP}). For the longitudinal spin densities (Eqs.~\eqref{eq:12} and \eqref{eq:14}), this does not occur because the terms responsible for the rotational behaviour are equally dependent on the gradient of the Laguerre polynomial. 

\section{Orbital angular momentum}

The (electric) orbital angular momentum density $\mathbf{l_E}$ can be obtained as~\cite{bliokh2013dual}
\begin{align}
\mathbf{l}_{\mathbf{E}} &= \frac{\epsilon_0}{2}\Im \mathbf{r} \times (\mathbf{E}^*\cdot\nabla\mathbf{E})\,,
\label{eq:17}
\end{align}
where it is understood that $(\mathbf{E}^*\cdot\nabla\mathbf{E})_j = \sum_{i=1}^{3}E^*_i\nabla_jE_i$ and the quantity $\Im\qty(\mathbf{E}^*\cdot\nabla\mathbf{E})$ is known as the electric canonical (orbital) momentum density $\mathbf{p_{\mathbf{E}}}^\text{o}$. 
Choosing a cylindrical coordinate frame, the $z$ component of the orbital angular momentum density in Eq.~\eqref{eq:17} is 
\begin{equation}\label{eq:18}
    l_{\mathbf{E}}^z = \frac{\varepsilon_0}{2}\,\Im p_{\mathbf{E}}^{\text{o},\phi}\,,
\end{equation}
so that the wavefront curvature can also generate a longitudinal orbital angular momentum density, as long as the canonical momentum has a nonzero azimuthal component. 
Using Eq.~\ref{eq:6} in Eqs.~\ref{eq:17},\ref{eq:18}, an expression for the general case can be obtained, but it is greatly simplified for a linearly polarized Gaussian beam ($\alpha = 1, \beta = 0, \ell = 0, p=0$). It is useful to split this calculation into the individual components with respect to the paraxial parameter:
\begin{align}
l_{{\mathbf{E}},z}^{\ell=p=0,\alpha=1} &= \frac{\epsilon_0}{2}\Im [(\mathbf{E}^{*\text{T0}}+\mathbf{E}^{*\text{L1}}  + \mathbf{E}^{*\text{T2}}  ) \nonumber \\ & \cdot \frac{\partial}{\partial\phi}(\mathbf{E}^{\text{T0}} + \mathbf{E}^{\text{L1}}  + \mathbf{E}^{\text{T2}})]\,.
\label{eq:19}
\end{align}
Calculations for $\ell=p=0$ reveal that $\frac{\partial}{\partial \phi}\mathbf{E}^{\text{T0}} = 0$ and  $\Im(\mathbf{E}^{*\text{L1}}\cdot\frac{\partial}{\partial \phi}\mathbf{E}^{\text{L1}}) = 0$, as well as $\Im(\mathbf{E}^{*\text{T2}}\cdot\frac{\partial}{\partial \phi}\mathbf{E}^{\text{T0}}) = 0$, leaving the following non-zero contribution:
\begin{align}
l_{{\mathbf{E}},z}^{\ell=p=0,\alpha=1} &= \frac{\epsilon_0}{2}\Im [\mathbf{E}^{*\text{T0}}\cdot \frac{\partial}{\partial\phi} \mathbf{E}^{\text{T2}}] \nonumber \\ 
&= \frac{4\epsilon_0 r^2}{kw^2\,R[z]}\sin\phi\cos\phi
|u_{\text{LG}}^{0,0}|^2\,.
\label{eq:20}
\end{align}
The resulting longitudinal orbital angular momentum density has the same magnitude but opposite sign of the spin angular momentum density Eq.~\eqref{eq:13}, so that they cancel each other out, yielding a null total angular momentum:
\begin{equation}\label{eq:21}
 \mathbf{j}_{\mathbf{E},z}^{\ell=p=0,\alpha=1} =\mathbf{s}_{\mathbf{E},z}^{\ell=p=0,\alpha=1} + \mathbf{l}_{\mathbf{E},z}^{\ell=p=0,\alpha=1} = 0\,.
\end{equation} 
Note that for beams with a nonzero degree of ellipticity in their 2D state of polarization ($\alpha,\beta\in\mathbb{C}$), $\Im(\mathbf{E}^{*\text{L1}}\cdot\frac{\partial}{\partial \phi}\mathbf{E}^{\text{L1}}) \neq 0$ and is responsible for the spin-to-orbit angular momentum conversion~\cite{zhao2007spin} in focused circularly-polarized Gaussian beams. As a final, intriguing point, the total angular momentum density in a more general case for $\ell\neq0$, $\mathbf{j}_{\mathbf{E},z}^{\ell, p, \alpha, \beta} \neq \mathbf{s}_{\mathbf{E},z}^{\ell, p, \alpha, \beta} + \mathbf{l}_{\mathbf{E},z}^{\ell, p, \alpha, \beta}$. This is to be expected given the fact that, for non-paraxial fields, separating the total angular momentum density into spin and orbital parts is problematic, though can be achieved~\cite{bliokh2014conservation}. 
\begin{figure*}[!ht]
    \centering
    \includegraphics[width = \linewidth]{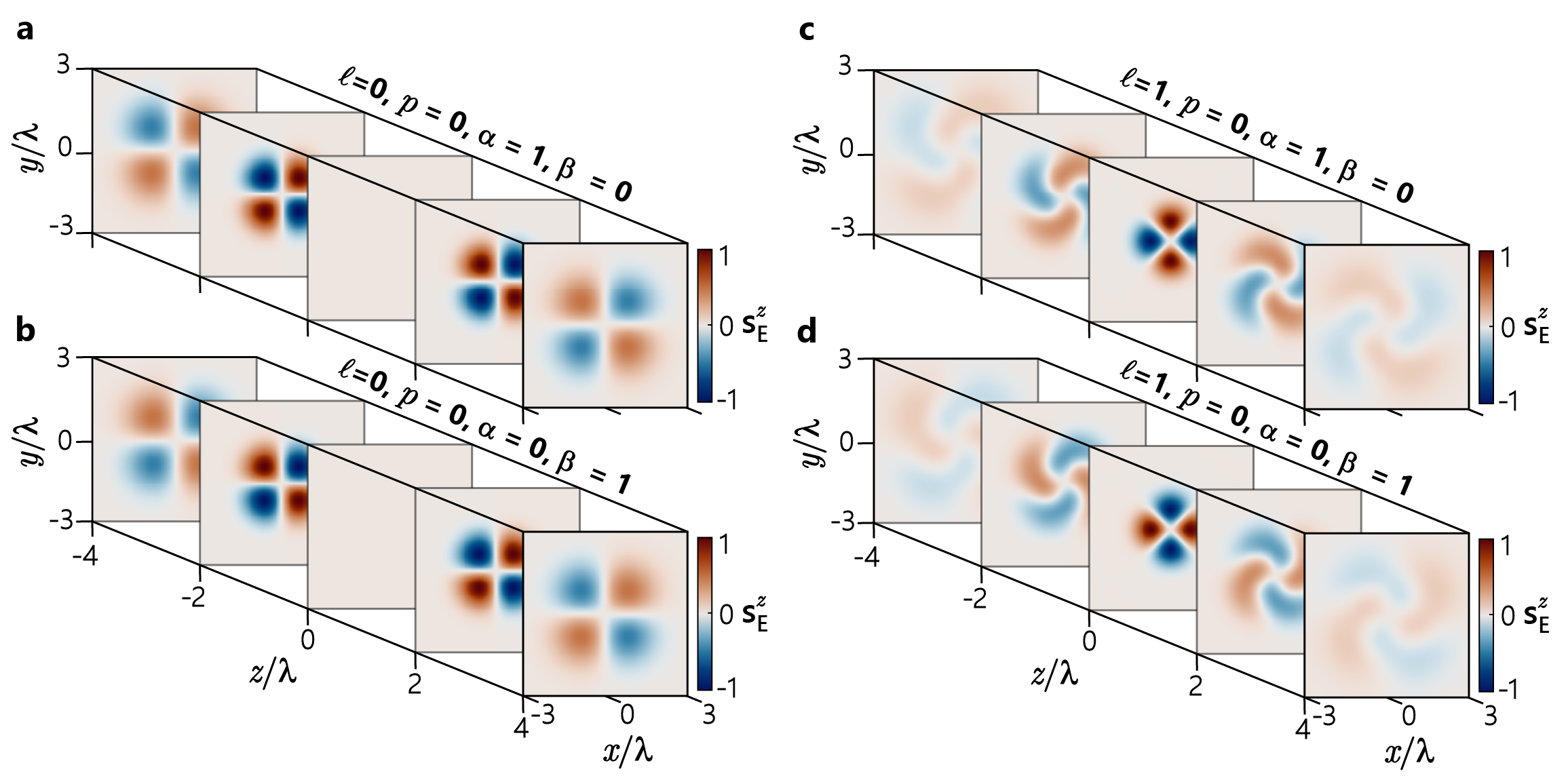}
    \caption{The same as shown in Fig.~\ref{fig:1}, calculated using the vectorial diffraction theory from the electric field defined in Eq.~\ref{eq:22}.}
    \label{fig:RW_fig1}
\end{figure*}

\section{Optical helicity}
It is generally assumed that optical helicity ($h$) and longitudinal spin angular momentum are associated with one another. It has recently been highlighted how such an assumption is in general not correct~\cite{forbes2024orbit}. In particular, the longitudinal spin angular momentum density of focused linearly polarized vortex beams (\emph{i.e.} Eq.~\eqref{eq:12} and Fig.\ref{fig:1}c,d) has zero optical helicity density associated with it~\cite{forbes2024orbit}. It is relatively straightforward to also confirm that the spin angular momentum of a linearly polarized Gaussian beam generated by the wavefront curvature (\emph{i.e.} Eqs.~\eqref{eq:13} and Fig.\ref{fig:1}a,b) likewise has zero optical helicity associated with it. The optical helicity density for a monochromatic beam is $h\propto-\Im\mathbf{E}^*\cdot\mathbf{B}$. Using the electric field (Eq.~\eqref{eq:6}) and the magnetic field (Eq.~\eqref{eq} in Supplementary~\ref{sec:Magn_field}), one can find $\mathbf{E}^{* \text{T0}}\cdot\mathbf{B}^{\text{T2}} = - \mathbf{E}^{* \text{T2}}\cdot\mathbf{B}^{\text{T0}}$ for 2D linearly polarized beams~\cite{forbes2021measures, forbes2024orbit} and, thus, $h=0$. Physically this means that whilst a focused linearly polarized Gaussian beam can produce torques $\mathbf{\tau}$ on absorbing particles with polarizability $\alpha$, via $\mathbf{\tau} \propto \Im(\alpha) \mathbf{s}_{\mathbf{E}}$~\cite{bliokh2014magnetoelectric}, it cannot produce differential absorption $A^{\text{ch}}$ in chiral particles due to having zero optical helicity density, as $A^{\text{ch}} \propto \Im(\alpha^\text{ch})h$~\cite{bliokh2014magnetoelectric}. Moreover, the dual-symmetric longitudinal spin angular momentum density, or longitudinal chiral spin angular momentum density~\cite{bliokh2014magnetoelectric}, of a focused linearly polarized LG beam (any optical vortex in fact~\cite{forbes2024orbit}) cannot exert a chiral radiation pressure force on chiral particles (see Supplementary~\ref{sec:Magn_field}).

\section{Semi-analytical modelling}
All the above properties can be confirmed numerically following the vectorial diffraction theory~\cite{richards1959, youngworth2000, novotny2012principles}. The chosen beam is decomposed in an infinite series of plane waves (the angular spectrum of the beam) which individually undergo refraction through a focusing element of a given numerical aperture (NA) and size (D). Such an approach has been recently implemented for a general vector vortex beam propagating through multilayered media~\cite{aita2024PRB}. Exploiting the open-source package therein proposed~\cite{Aita2024code}, the focusing of a general scalar vortex described by Eqs.~\ref{eq:1} can be modelled for free space propagation. The resulting electric field can be written as
\begin{align}
    \vb{E}_{\ell p}\qty(\vb{r}) &= \mathrm{C}_{\ell p}\bigintss_0^{\theta_{\text{max}}}\hspace{-0.6cm}\dd\theta\,f(\theta)\, \sqrt{\cos\theta}\sin\theta\mqty(\tilde{E}_x[\theta;\vb{r}]\\\tilde{E}_y[\theta;\vb{r}]\\\tilde{E}_z[\theta;\vb{r}])\, \text{e}^{ikz\cos\theta}\label{eq:22}\\
    \text{with:}& \nonumber\\
    \mathrm{C}_{\ell p}&= -i k f \text{e}^{-i k f}\,  \sqrt{\frac{\pi 2^{|\ell|} p!}{2(| \ell | +p)!}}\label{eq:23}\\
    f(\theta) &= L_p^{|\ell|}\qty[\frac{2f^2\sin^2\theta}{w_0^2}]\qty(\frac{f\sin\theta}{w_0})^{|\ell|} \text{e}^{-\frac{f^2\sin^2\theta}{w_0^2}} \label{eq:24}
\end{align}
where the integration limits are given by the angular aperture of the focusing element (NA = $\sin\theta_{\text{max}}$). The components of the integrand vector $\tilde{E}_i$ are obtained after an analytical integration over the in-plane angle $\phi$, resulting in combinations of several orders of Bessel functions of the first kind ($J_\ell[\Psi]$). In the simpler case of a Gaussian beam, obtained for $\ell=p=0$ of a generic state of polarization given by Eq.~\ref{eq:1}, the $\tilde{E}_i$ components are
\begin{subequations}\label{eq:25}
\begin{align}
    &\tilde{E}_x = \nonumber\\
    &\alpha J_0[\Psi] \qty(1+\cos\theta) + 2J_2[\Psi]\sin^2\frac{\theta}{2} \qty(\alpha\cos2\phi + \beta\sin2\phi)\\
    &\tilde{E}_y = \nonumber\\
    & \beta J_0[\Psi]\qty(1+\cos\theta) + 2J_2[\Psi]\sin^2\frac{\theta}{2}\qty(\alpha\sin2\phi - \beta\cos2\phi)\\
    &\tilde{E}_z = \nonumber\\
    &-2i J_1[\Psi]\sin\theta \qty(\alpha \cos\phi + \beta \sin\phi)\,, 
\end{align}
\end{subequations}
where the argument of each Bessel function is given by $\Psi(\theta) = kr\sin\theta$. Substituting the electric field obtained with Eq.~\ref{eq:22} into Eqs.~\ref{eq:8}, the same longitudinal and transversal spin densities calculated with the fully analytical approach can be obtained (cf. Fig.~\ref{fig:1},~\ref{fig:RW_fig1} and  Supplementary Figs.~\ref{fig:si_RW_circ} and ~\ref{fig:si_RW_tc}).

\section{Discussion and Conclusion}
The transverse spin momentum of focussed beams is a well-known phenomenon~\cite{bliokh2015transverse}, in part because it is proportional to the paraxial parameter to the first-order and also because its components generally act in directions which have no competing optical angular momenta produced by the dominant zeroth-order fields. In contrast, the longitudinal spin angular momentum (Eq.~\eqref{eq:11}) is proportional to the paraxial parameter to the second-order so its generation requires a strongly focused beam. While for elliptically polarized beams the contribution of the wavefront curvature to the longitudinal spin is overshadowed by a conventional longitudinal spin of the zeroth order, for 2D linearly-polarized beams, a conventional longitudinal spin is absent and we have shown a linearly polarized Gaussian beam possesses a nonzero longitudinal spin solely due to the wavefront curvature. Other momenta at sub-wavelength scale and the subsequent torques and forces they subject material probes to have benefited from such unique characteristics in their experimental observation, e.g. the Belinfante spin momentum~\cite{antognozzi2016direct}.

It is interesting and important to reflect upon the approximations made in the analytical theory developed above. Namely, neglecting the $z$-variations in the amplitude distribution, which leads to Eq.~\eqref{eq:5}, and using paraxial expression for LG beams as a starting point to describe strongly non-paraxial scenarios. Comparing the analytical results with the vectorial diffraction theory, the correctness of the presented analytical approach has been validated. The analytical analysis has allowed us to determine the important and unique contributions of the gradient of wavefront curvature to generation of angular momenta--insight which is hidden in the numerical analysis. Moreover, it is important to also make the point that in order to unearth the novel contributions to the longitudinal spin and orbital angular momentum, we had to derive the analytical electromagnetic fields up to second-order in the paraxial parameter, i.e. the phenomena stem from the second-order transverse field components. Whilst the first-order longitudinal fields are often incorporated in analytical theories of focused beams~\cite{adams2018optics}~, the second-order transverse components are not. By including them in this work, we have unearthed novel contributions to longitudinal optical angular momenta and highlighted their necessity in studies of angular momenta of focused beams. 

Whilst inherently interesting phenomena, the optical angular momentum of focused beams discussed above also offers new methods for the development of optical manipulation and light-matter interactions. The results show that beams with zero angular momentum (both spin and orbital) in the far field, once strongly focused have non-zero spin and orbital angular momentum in the focal region. In particular, they can, therefore, generate torques $\mathbf{\tau} \propto \Im(\alpha) \mathbf{s}_{\mathbf{E}}$, where $\alpha$ is the electric polarizability, in probe particles from light which in the far field has zero spin angular momentum. In principle, it thus allows for spinning particles with just a high-NA lens. Moreover, $\mathbf{s}_{\mathbf{E}}$ is responsible for magnetic circular dichroism, with a similar interaction occurring in light-atom transitions in an external static magnetic field~\cite{bliokh2015transverse}. The fact $\mathbf{s}_{\mathbf{E}}$ can be non-zero for 2D linearly polarized light immediately, therefore, suggests the ability to probe matter through `magnetic linear dichroism'. 

Another interesting avenue for future exploration would be to study the polarization topology which produces the novel spin angular momenta described above. As mentioned in Section III, Eq.~\eqref{eq:8} is simply a measure of the phase difference between two orthogonal electric field components, which yield a maximum value of $\pi/2$ for circularly polarized light. Simple inspection of Eq.~\eqref{eq:6} clearly shows that, for $\alpha = 1$ and $\ell=p=0$ for example, the T0 field polarized in $\mathbf{\hat{x}}$ is $\pi/2$ out-of-phase with the T2 $\mathbf{\hat{y}}$ component $\sin2\phi\Im\gamma^2/2k^2 = -2\sin2\phi r^2/R[z]kw^2$. Therefore in the $x,y$ plane the electric field vector traces out an elliptical (in general) path, generating the local electric spin angular momentum density (Figure~\ref{fig:1}). Under paraxial conditions, the T2 field is essentially non-existent because $1/kw\ll1$. The paraxial beam is thus essentially just $x$-polarized in the $x,y$ plane due to the dominant T0 field and has no spin angular momentum density; as the beam becomes more focused, the ratio of $\lambda/w$ grows and the polarization vector becomes elliptical (with spatial variation of course) in the $x,y$-plane due to the $\pi/2$ out-of-phase (imaginary) T2 $y$-component which originates from the gradient of the wavefront curvature. Analogous analysis with recourse to the polarization properties of the electric field applies to the other spin angular momentum densities studied here. 

In summary, we have highlighted how counter-intuitive angular momentum properties of light manifest at the subwavelength scales: focused linearly polarized Gaussian beams can possess spin and orbital angular momentum around the focal plane (but not at $z=0$) and focused circularly polarized beams have a helicity-dependent transverse spin angular momentum. Taking advantage of the full analytical approach, the physical origin of both of these effects has been found to be in the gradient of the wavefront curvature. Numerical calculations based on the commonly used vectorial diffraction theory have also been performed, showing an extremely good match with analytical results. This work has highlighted the strong potential for the wavefront curvature to be utilized as a degree of freedom of light for applications in nano-optics. 

%The physical origin of both of these effects is in the gradient of the wavefront curvature.
%We used analytical theory to discover the physical origins of these angular momenta, and verified our theory with numerical modelling based on Richards-Wolf diffraction theory. This work has highlighted the strong potential for the wavefront curvature to be utilized as a degree of freedom of light for applications in nano-optics. 

\section*{Funding}
V.A. and A.V.Z. acknowledge support from the ERC iCOMM project (789340) and the UK EPSRC project EP/Y015673/1.
All the data supporting the findings of this work are presented in the results sections and available from the corresponding author upon reasonable request.

\section*{Disclosures}
The authors declare no competing interests.

\section*{Data availability} No data were generated or analyzed in the presented research.

\newpage
\bibliography{references.bib}

%%%%%%%%%% Embedding supplemental materials %%%%%%%%%%
\onecolumngrid
\clearpage

\begin{center}
\textbf{\large Supplemental Materials: Generating optical angular momentum through wavefront curvature}
\end{center}

%%%%%%%%%% Prefix a "S" to all equations, figures, tables and reset the counter %%%%%%%%%%
\setcounter{equation}{0}
\setcounter{figure}{0}
\setcounter{table}{0}
\setcounter{section}{0}
\setcounter{page}{1}
\makeatletter
\renewcommand{\theequation}{S\arabic{equation}}
\renewcommand{\thefigure}{S\arabic{figure}}
\renewcommand{\bibnumfmt}[1]{[S#1]}
\renewcommand{\citenumfont}[1]{S#1}
%%%%%%%%%% Prefix a "S" to all equations, figures, tables and reset the counter %%%%%%%%%%

\section{Magnetic Field Derivation}\label{sec:Magn_field}

The magnetic field of a Laguerre-Gaussian mode truncated at second-order in the paraxial parameter is

\begin{align}
    \mathbf{B}^\text{T0+L1+T2}_{\text{LG}} &= \Biggl\{ \overbrace{\alpha \mathbf{\hat{y}} - \beta \mathbf{\hat{x}}}^{\text{T0}} + \overbrace{\mathbf{\hat{z}}\frac{i}{k} \Biggl[\alpha \qty( \gamma \sin\phi + \frac{i\ell}{r} \cos\phi ) - \beta \qty( \gamma \cos\phi - \frac{i\ell}{r} \sin\phi ) \Biggr]}^{\text{L1}} \nonumber \\
    &+ \frac{1}{k^2} \mathbf{\hat{x}} \Biggl[\alpha\sin\phi\cos\phi\qty(\gamma^{'} + \gamma^2 - \frac{\gamma}{r} + \frac{\ell^2}{r^2}) + \alpha\cos^2\phi\qty(\frac{i\ell\gamma}{r}-\frac{i\ell}{r^2})+\alpha\sin^2\phi\qty(\frac{i\ell}{r^2} - \frac{i\ell\gamma}{r}) \nonumber \\ & + 2\beta\sin\phi\cos\phi\qty(\frac{i\ell\gamma}{r}-\frac{i\ell}{r^2} ) + \beta \cos^2\phi \qty(\frac{\gamma}{r} - \frac{\ell^2}{r^2} ) + \beta \sin^2\phi \qty(\gamma^{'} + \gamma^2 ) \Biggr]\nonumber \\
    &- \frac{1}{k^2} \mathbf{\hat{y}} \Biggl[2\alpha\sin\phi\cos\phi\qty(\frac{i\ell}{r^2}-\frac{i\ell\gamma}{r}) + \alpha\cos^2\phi\qty(\gamma^{'} + \gamma^{2}) + \alpha\sin^2\phi\qty(\frac{\gamma}{r}-\frac{\ell^2}{r^2}) \nonumber \\
    &+ \beta\sin\phi\cos\phi\qty(\gamma^{'} + \gamma^2 - \frac{\gamma}{r} + \frac{\ell^2}{r^2} ) + \beta\cos^2\phi\qty(\frac{i\ell\gamma}{r} - \frac{i\ell}{r^2}) + \beta\sin^2\phi\qty(\frac{i\ell}{r^2}-\frac{i\ell\gamma}{r})\Biggr]\Biggr\} \frac{ u_{\text{LG}}^{\ell,p}}{c}.
    \label{eq}
\end{align}

With an analogous calculation to that carried in Section III of the main manuscript, i.e. inserting Eq.~\eqref{eq} into $\mathbf{s_B} = \frac{c^2\epsilon_0}{2} \Im\mathbf{B}^*\times\mathbf{B}$, the magnetic spin angular momentum density for a Laguerre-Gaussian beam can be truncated to the second order of the paraxial parameter. The zeroth-order contribution comes from the terms T0$\times$T0, the first-order ones from T0$\times$L1 and L1$\times$T0, the second-order from T0$\times$T2 and T2$\times$T0:

\begin{align}
\mathbf{s_B}^{\text{T0} \times \text{T0}} &= \frac{c^2\epsilon_0}{2} \Im \bigl(\mathbf{B}^{*\text{T0}}_\text{LG} \times \mathbf{B}^{\text{T0}}_\text{LG}\bigr) \nonumber \\ 
& = \mathbf{\hat{z}}\frac{\epsilon_0}{2}\Im\bigl(\alpha^{*}\beta - \alpha \beta^{*}\bigr)|u_{\text{LG}}^{\ell,p}|^2\,, 
\label{eq1}\\
\mathbf{s_B}^{\text{T0} \times \text{L1}} &= \frac{c^2\epsilon_0}{2} \Im \bigl(\mathbf{B}^{*\text{T0}}_\text{LG} \times \mathbf{B}^{\text{L1}}_\text{LG} + \mathbf{B}^{*\text{L1}}_\text{LG} \times \mathbf{B}^{\text{T0}}_\text{LG} \bigr) \nonumber \\ 
& = \Im\frac{i\epsilon_0}{2k}\Bigl\{ \mathbf{\hat{x}} \Bigl[|\alpha|^2 \qty(\gamma\sin\phi+\frac{i\ell}{r}\cos\phi) - \alpha^*\beta\qty(\gamma\cos\phi-\frac{i\ell}{r}\sin\phi) + \text{c.c.}\Bigr] \nonumber \\ 
& + \mathbf{\hat{y}} \Bigl[\alpha\beta^* \qty(\gamma\sin\phi+\frac{i\ell}{r}\cos\phi) - |\beta|^2\qty(\gamma\cos\phi-\frac{i\ell}{r}\sin\phi) + \text{c.c.}\Bigr] \Bigr\}
|u_{\text{LG}}^{\ell,p}|^2\,, 
\label{eq11}\\
\mathbf{s_B}^{\text{T0} \times \text{T2}} &= \frac{c^2\epsilon_0}{2} \Im \bigl(\mathbf{B}^{*\text{T0}}_\text{LG} \times \mathbf{B}^{\text{T2}}_\text{LG} + \mathbf{B}^{*\text{T2}}_\text{LG} \times \mathbf{B}^{\text{T0}}_\text{LG} \bigr) \nonumber \\
& = \mathbf{\hat{z}}\Im\frac{\epsilon_0}{2k^2}\Bigl\{ \Bigl[-|\alpha^2|\sin\phi\cos\phi\qty(\gamma^{'}+\gamma^2-\frac{\gamma}{r}+\frac{\ell^2}{r^2}) - 2\alpha^*\beta\sin\phi\cos\phi\qty(\frac{i\ell\gamma}{r}-\frac{i\ell}{r^2}) - |\alpha^2|\cos^2\phi\qty(\frac{i\ell\gamma}{r}-\frac{i\ell}{r^2})  \nonumber \\ 
& - |\alpha^2|\sin^2\phi \Bigr\{\frac{i\ell}{r^2}-\frac{i\ell\gamma}{r}) -\alpha^*\beta\sin^2\phi\qty(\gamma^{'}+\gamma^2) - \alpha^*\beta\cos^2\phi\qty(\frac{\ell^2}{r^2}-\frac{\gamma}{r}) -\text{c.c.} \Bigr] \nonumber \\  
& + \Bigr[2\alpha\beta^*\sin\phi\cos\phi\qty(\frac{i\ell}{r^2}-\frac{i\ell\gamma}{r}) - \alpha\beta^*\sin^2\phi\qty(\frac{\ell^2}{r^2}-\frac{\gamma}{r})+\alpha\beta^*\cos^2\phi\qty(\gamma^{'}+\gamma^2) \nonumber \\ 
& + |\beta^2|\sin\phi\cos\phi\qty(\gamma^{'}+\gamma^2-\frac{\gamma}{r}+\frac{\ell^2}{r^2}) + |\beta^2|\cos^2\phi\qty(\frac{i\ell\gamma}{r}-\frac{i\ell}{r^2}) + |\beta^2|\sin^2\phi\qty(\frac{i\ell}{r^2}-\frac{i\ell\gamma}{r}) - \text{c.c.} \Bigr] \Bigr\}
|u_{\text{LG}}^{\ell,p}|^2\,,  
\label{eq111}
\end{align}

Note that the paraxial contribution (pure zeroth-order fields) to the magnetic spin (Eq.~\eqref{eq1}) is identical to the paraxial contribution to the electric spin (Eq.~\eqref{eq:9}), but the spin angular momentum densities which depend on non-paraxial electric and magnetic fields ( Eqs.~\eqref{eq:10} and \eqref{eq:11} and Eqs.~\eqref{eq11} and \eqref{eq111}) are dual-asymmetric with respect to one another. The electric spin angular momentum densities in the main manuscript (Eqs.~\eqref{eq:9}--\eqref{eq:11}) when added to their magnetic counterparts (Eqs.\eqref{eq1}--\eqref{eq111}) yield the dual-symmetric spin angular momentum density $\mathbf{s} = \mathbf{s_\mathbf{E}} + \mathbf{s_\mathbf{B}}$. This dual-symmetric spin angular momentum density contributes to a chiral radiation pressure force on chiral particles~\cite{bliokh2014magnetoelectric}. As mentioned in Section V, for 2D linearly polarized beams, the dual-symmetric longitudinal spin is identically zero due to the fact that $\mathbf{s_E}^{\text{T0} \times \text{T2}} = - \mathbf{s_B}^{\text{T0} \times \text{T2}}$, thus cannot produce a chiral radiation pressure force~\cite{forbes2024orbit}.

\section{Additional examples from the full analytical model}
\begin{figure}[ht]
    \centering
    \includegraphics[width=\linewidth]{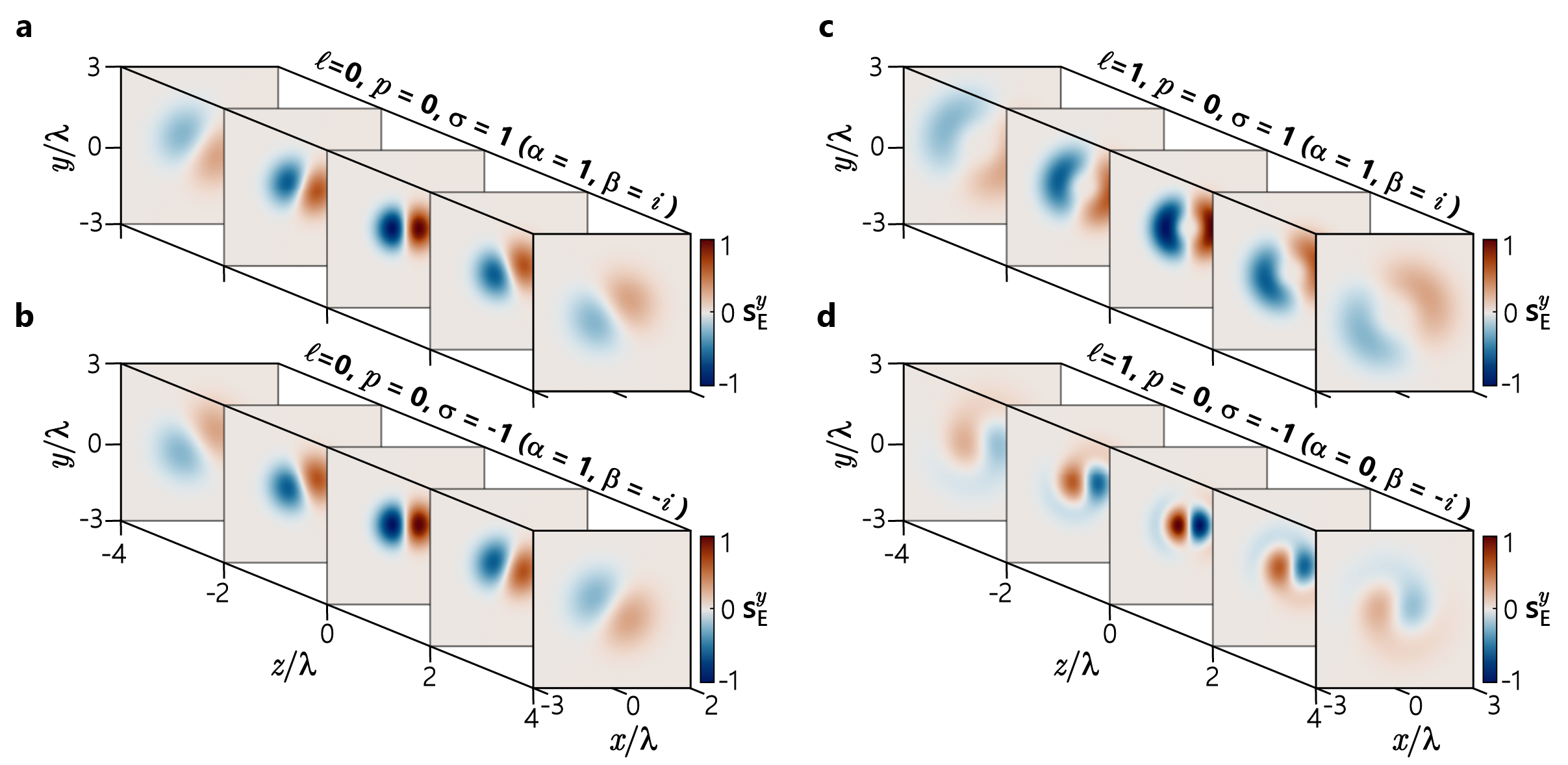}
    \caption{The dependence of the transverse spin density $\mathbf{s_E}^{y}$ component on a propgation distance $z$, for an LG beam with $p=0$, (a,b) $\ell$ =0, (c,d) $\ell$=1 and circular polarization of (a,c) left- and (b,d) right- handedness. Corrsponding $\mathbf{s_E}^{x}$ dependences are shown in Fig.~\ref{fig:trans_comps_lg00_circ}c,d and Fig.~\ref{fig:x_comp_lg10_circ}.}
    \label{fig:si_ycomps}
\end{figure}

\begin{figure}[ht]
    \centering
    \includegraphics[width=\linewidth]{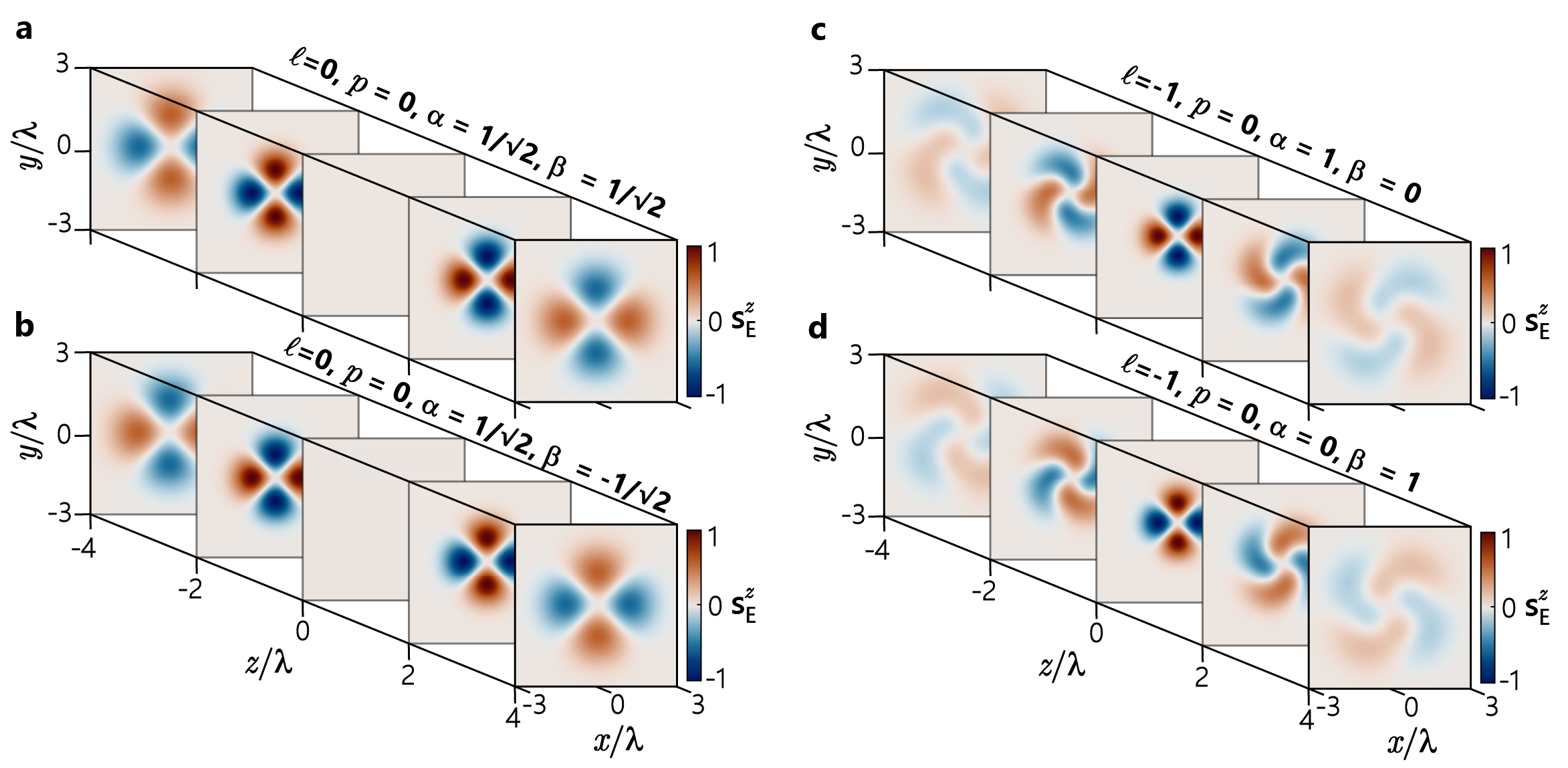}
    \caption{The dependence of the longitudinal spin density $\mathbf{s_E}^{z}$ on the propgation distance $z$ for LG beams of orders (a,b) $\ell=p=0$, (c,d) $\ell = -1,\,p=0$ and (a) diagonal ($\alpha = \beta = 1/\sqrt{2}$), (b) antidiagonal ($\alpha = -\beta = 1/\sqrt{2}$), (c) horizontal ($\alpha = 1,\beta = 0$) and (d) vertical ($\alpha = 0,\beta = 1$) state of polarization.}
    \label{fig:si_extra_cases}
\end{figure}

\begin{figure}[ht]
    \centering
    \includegraphics[width=\linewidth]{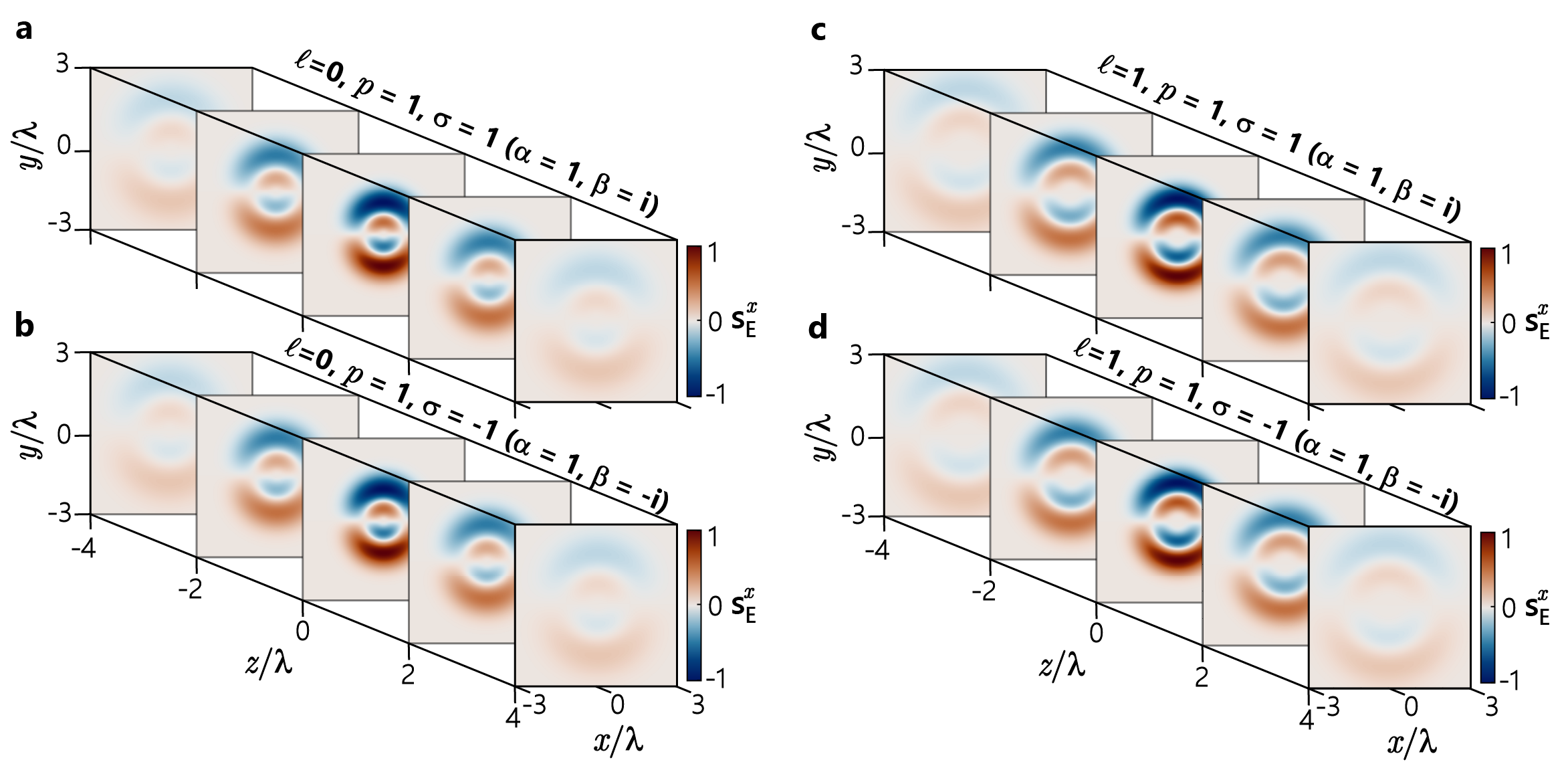}
    \caption{The dependence of the transverse spin density $\mathbf{s_E}^{x}$ component on ta propagation distsnce $z$ for LG beams of nonzero radial order ($p=1$) and (a,b) $\ell = 0$, (c,d) $\ell = 1$ and circular polarization with (a,c) right- and (b,d) left- handedness.}
    \label{fig:si_nonzeroP}
\end{figure}

\section{Additional examples from the vectorial diffraction theory}

For the general case of a Laguerre-Gauss beam with $(\ell,p)\neq0$, the components of the integrated vector in Eq.~\ref{eq:22} are given by

 \begin{align}
 \tilde{E}_x &= -i^{-\ell } (\alpha-i \beta) (\cos\theta-1) e^{-i (\ell -2) \phi } J_{2-\ell }[\Psi]-i^{\ell } (\alpha+i \beta) (\cos\theta-1) e^{-i (\ell +2) \phi } J_{\ell +2}[\Psi]\nonumber\\
 &+2 \alpha i^{\ell } (\cos\theta+1) e^{-i \ell  \phi } J_{\ell }[\Psi]\\
\tilde{E}_y &=i^{-\ell } (\beta+i \alpha) (\cos\theta-1) e^{-i (\ell -2) \phi } J_{2-\ell }[\Psi]+i^{\ell } (\beta-i \alpha) (\cos\theta-1) e^{-i (\ell +2) \phi } J_{\ell +2}[\Psi]\nonumber\\
&+2 \beta i^{\ell } (\cos\theta+1) e^{-i \ell  \phi } J_{\ell }[\Psi]\\
\tilde{E}_z &=2 \sin\theta e^{-\frac{1}{2} i \ell  (2 \phi +\pi )} \qty[(\beta-i \alpha) e^{i (\pi  \ell -\phi )} J_{\ell +1}[\Psi]+(\alpha-i \beta) (\sin (\phi )-i \cos (\phi )) J_{1-\ell }[\Psi]]\, .
\end{align}   

\begin{figure}[h]
    \centering
    \includegraphics[width = \linewidth]{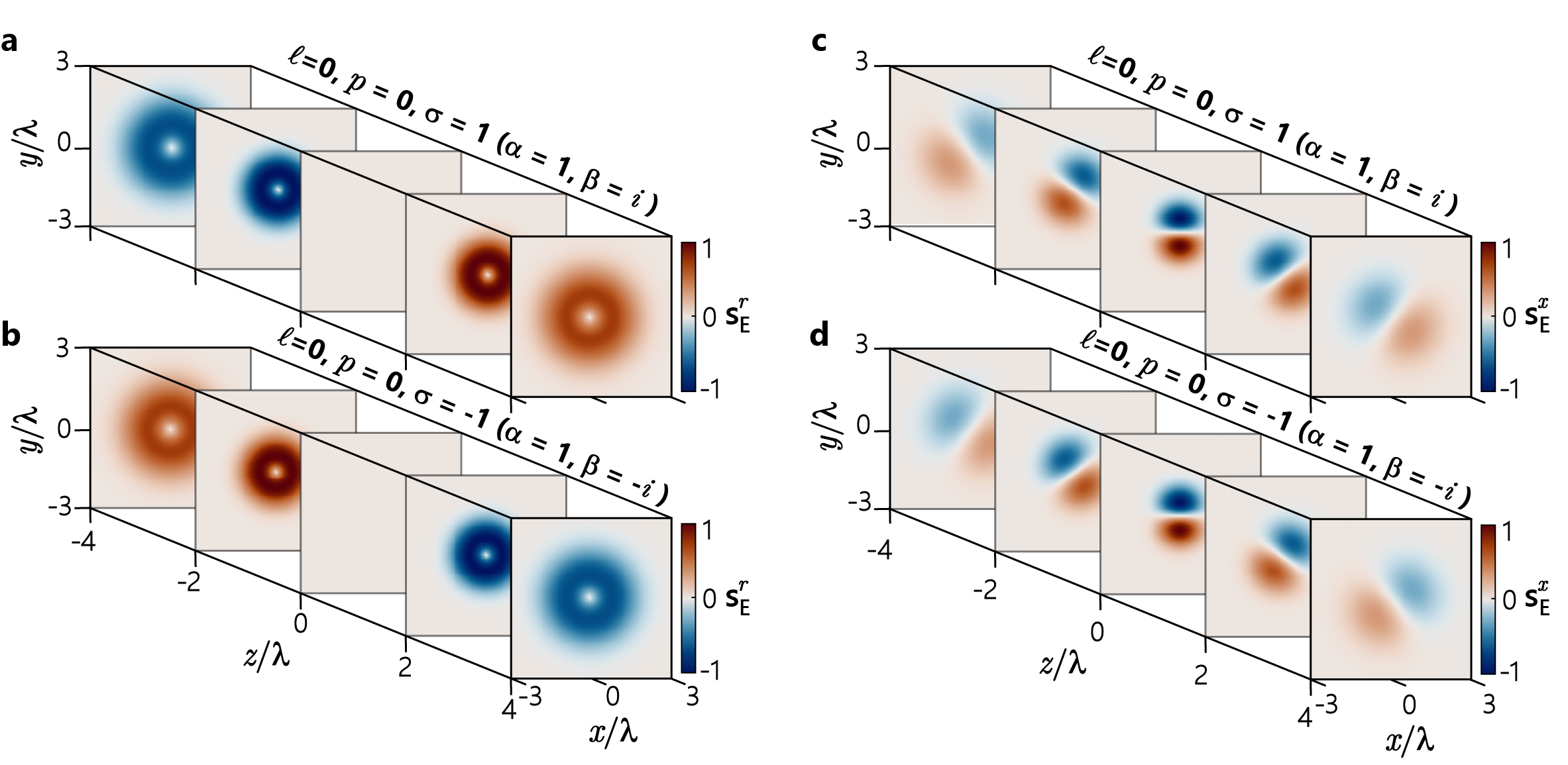}
    \caption{The same as shown in Fig.~\ref{fig:trans_comps_lg00_circ}, calculated with the vectorial diffraction theory from the electric field defined in Eq.~\ref{eq:22} choosing input parameters matching the system described with the analytical theory.}
    \label{fig:si_RW_circ}
\end{figure}

\begin{figure}
    \centering
    \includegraphics[width = 0.5\linewidth]{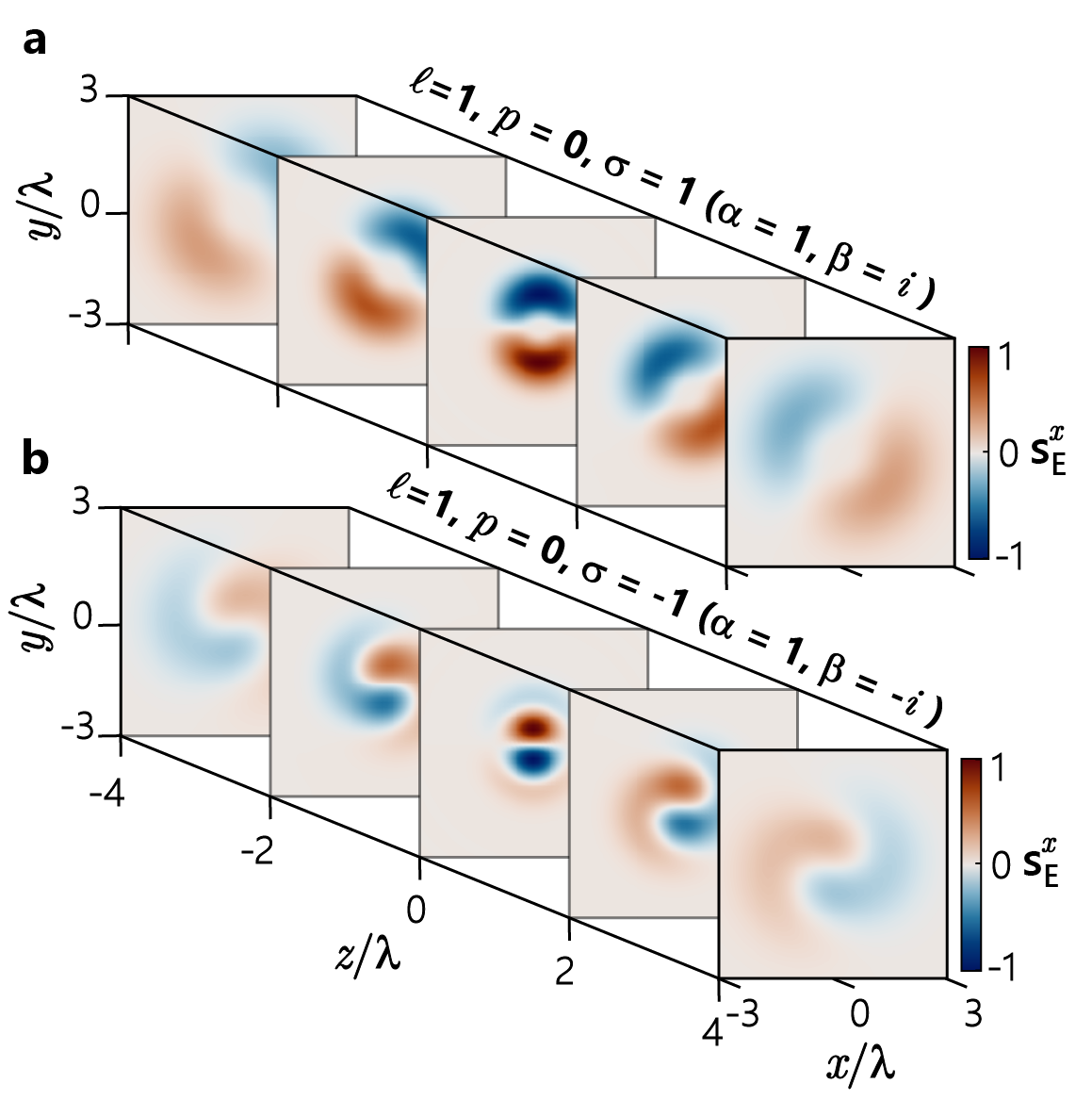}
    \caption{The same as shown in Fig.~\ref{fig:x_comp_lg10_circ}, calculated with the vectorial diffraction theory from the electric field defined in Eq.~\ref{eq:22}.}
    \label{fig:si_RW_tc}
\end{figure}

Upon integration of these expressions, the obtained electric field can be used in Eq.~\ref{eq:8} to achieve the same results as the full-analytical model (Fig.~\ref{fig:si_RW_circ}--\ref{fig:si_RW_tc}).

\begin{figure}
    \centering
    \includegraphics[width = 0.5\linewidth]{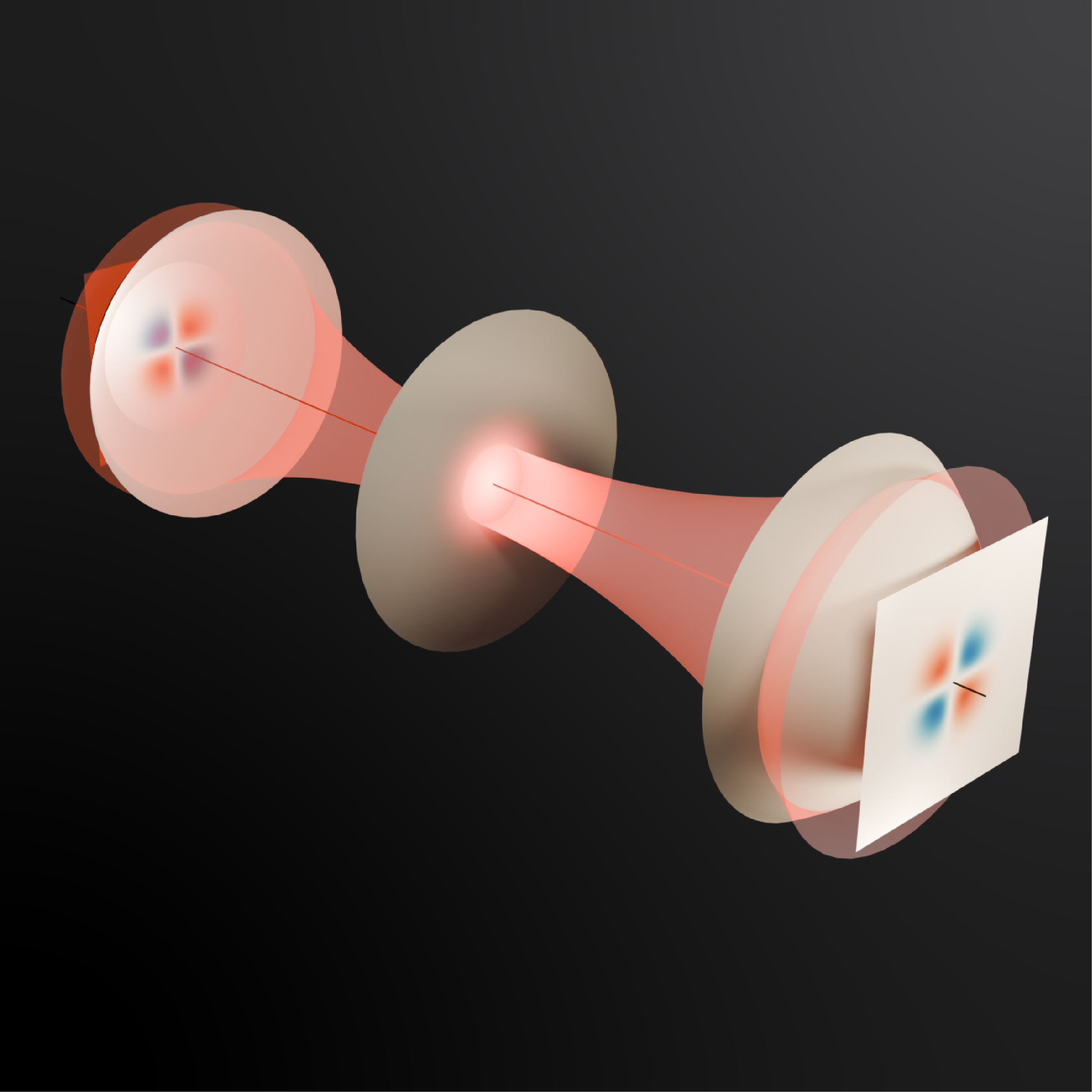}
    \caption{Table of Contents image.}
    \label{fig:TOC}
    \end{figure}

\end{document}